\documentclass[aps,prb,twocolumn,floatfix,superscriptaddress,amssymb,amsmath]{revtex4}
\usepackage{graphicx}

\begin{document}

\title{Coherent control of tunneling in a quantum dot molecule}
\author{J.~M.~Villas-B\^{o}as}
\affiliation{Departamento de F\'{\i}sica, Universidade Federal de
S\~{a}o Carlos, 13565-905, S\~{a}o Carlos, S\~{a}o Paulo, Brazil}
\affiliation{Department of Physics and Astronomy, Condensed Matter
and Surface Science Program, \\Ohio University, Athens, Ohio
45701-2979}
\author{A.~O.~Govorov}
\affiliation{Department of Physics and Astronomy, Condensed Matter
and Surface Science Program, \\Ohio University, Athens, Ohio
45701-2979}
\author{Sergio E.~Ulloa}
\affiliation{Department of Physics and Astronomy, Condensed Matter
and Surface Science Program, \\Ohio University, Athens, Ohio
45701-2979}

\begin{abstract}
We demonstrate theoretically that it is possible to use Rabi
oscillations to coherently control the electron tunneling in an
asymmetric double quantum dot system, a quantum dot molecule. By
applying an optical pump pulse we can excite an electron in one of
the dots, which can in turn tunnel to the second dot, as
controlled by an external voltage. Varying the intensity of the
pulse one can suppress or enhance the tunneling between the dots
for given level resonance conditions.  This approach allows
substantial flexibility in the control of the quantum mechanical
state of the system.

\end{abstract}

 \maketitle

Quantum dot ({\bf QD}) structures provide a three-dimensional
confinement of carriers. Electrons and holes in the QD can occupy
only a set of states with discrete energies, just as in an atom,
and can thus be used to perform ``atomic physics" experiments in
solid state structures. One advantage of QDs is that they provide
different energy scales and physical features which can be easily
varied over a wide range of values. Most important, perhaps, is
that QDs also allow the control of direct quantum mechanical
electronic coupling with not only composition but externally
applied voltages. These flexible systems represent therefore the
ideal for theoretical and experimental investigations, where the
interactions between light and matter can be studied in a fully
controlled, well-characterized environment, and with excellent
optical and electrical probes. These features make semiconductor
QDs promising candidates for applications in electro-optical
devices such as QD lasers, \cite{Saito99,Shchekin00} and in
quantum information
processing.\cite{Ekert96,Sherwin99,Loss98,Imamoglu99} In the
latter case, one can exploit the optical excitation in a QD,
\cite{Ekert96,Sherwin99} or its spin state,
\cite{Loss98,Imamoglu99} as qubits. These high expectations are
produced by experimental advances in the coherent manipulation of
QD states, such as the exciton Rabi oscillations in single dots,
achieved by the application of electromagnetic
pulses.\cite{Stievater01,Kamada01,Htoon02,Zrenner02,Li03} Coherent
phenomena in ensembles of QDs have also been
observed.\cite{Cole01,Borri02,Michler00,Kim99,Pelton02} Similarly,
lithographically-defined dots have shown great potential in the
control of coherently coupled
systems.\cite{Blick98,Oosterkamp98,Holleitner02}

The ability to assemble collections of QDs with designed
geometries opens up a number of interesting possibilities. Here we
describe theoretically the behavior of a QD-molecule formed from
an asymmetric double QD system coupled by tunneling. Such a QD
molecule can be fabricated using self-assembled dot growth
technology.\cite{Petroff01} By applying a near resonant optical
pulse we can excite one electron from the valence to the
conduction band in one dot, which can in turn tunnel to the second
dot. We show that by suitably varying the frequency detuning or
applied voltage on the QD pair, it is possible to control the
inter-dot oscillations in the system, or use them to monitor the
coherent state of one dot in the presence of the radiation field.
This opens up the possibility of controlling the quantum
mechanical state of such structure, perhaps useful in the field of
quantum computation and information storage.

\begin{figure}[tbp]
\includegraphics*[width=1.0\linewidth]{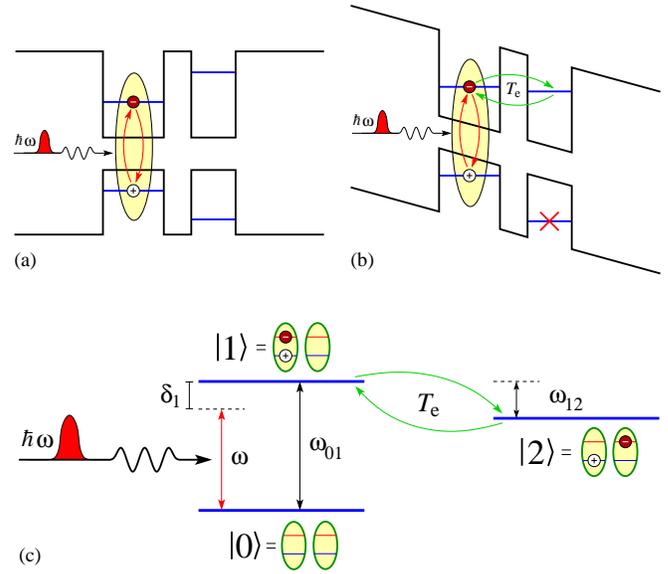}
\caption{(Color online) Schematic band structure and level
configuration of a double QD system. (a) Without a gate voltage,
electron tunneling is weak. (b) With applied gate voltage,
conduction band levels get into resonance, increasing their
coupling, while valence band levels become even more
off-resonance, resulting in effective decoupling of those levels.
(c) Levels taken into account by Hamiltonian model and the basis
of states used to describe it. A pulsed laser excites one electron
from the valence band that can tunnel to the other dot. We assume
that the hole cannot tunnel in the time scale we are considering
here.} \label{schematic}
\end{figure}

A schematic representation of the Hamiltonian for our model can be
seen in Fig.\ \ref{schematic}. In Fig.\ \ref{schematic}(a),
without a gate voltage, the levels are out of resonance, which
results in weak interdot tunneling.  In contrast, Fig.\
\ref{schematic}(b) shows the situation with a gate voltage, where
the conduction band levels get closer to resonance, greatly
increasing their coupling, while the valence band levels get more
off-resonance, resulting in effective decoupling of those levels.
Considering this last configuration, we can neglect the hole
tunneling and write the Hamiltonian as represented in Fig.\
\ref{schematic}(c). The electromagnetic field is introduced by the
usual dipole interaction, which couples the ground state
$|0\rangle$ (the system without excitations) with the exciton
state $|1\rangle$ (a pair of electron and hole bound in the first
dot).\footnote{The biexciton state is far off resonance due the
electron-electron interaction and can safely be neglected in our
analysis for long pulses (requiring low intensity to achieve a
$\pi$-pulse), as has been experimentally demonstrated for a single
dot.\cite{Stievater01,Kamada01,Htoon02,Zrenner02,Li03}} The
electron tunneling, on the other hand, couples the exciton
$|1\rangle$ with the indirect exciton state $|2\rangle$ (one hole
in the first dot with an electron in the second dot). Using this
configuration the Hamiltonian can be written as \footnote{Assuming
a left-polarized pulse or linearly polarized light in rotation
wave approximation (RWA), which can be applied here since the
pulse intensity is much smaller than the separation of levels
coupled and frequency of the pulse.}

\begin{eqnarray}
H &=&\sum_{j}\varepsilon_{j}|j\rangle \langle j|+T_{e}\left(
{|1\rangle \langle 2|+|2\rangle \langle 1|}\right) \nonumber \\
&&+\hbar\Omega\left( {e^{-i\omega t}| 0\rangle \langle 1|
+e^{i\omega t}|1\rangle \langle 0|}\right), \label{t1}
\end{eqnarray}
where $\varepsilon_{j}$ is the energy of the state $|j\rangle$,
$T_{e}$ is the electron tunneling matrix element, and
$\Omega(t)=\langle 0| \vec{\mu}\cdot \vec{E}(t)|1\rangle/2\hbar$,
where $\vec{\mu}$, the electric dipole moment, describes the
coupling to the radiation field of the excitonic transition, and
$\vec{E}(t)$ is the pulse amplitude which can have different
shapes. \footnote{Notice that direct coupling from $|0\rangle$ to
$|2\rangle$ is neglected, as the dipole moment for that spatially
indirect exciton will be vanishingly small.}

To simplify the time-dependent Schr\"{o}dinger equation, we
utilize the unitary transformation
\begin{equation}
|\psi \rangle = \exp \left[ -\frac{i\omega t}{2}\left(|1\rangle
\langle 1| -|0\rangle \langle 0| +|2\rangle \langle 2|\right)
\right] |\psi^{\prime}\rangle,\label{t2}
\end{equation}
which removes the fast oscillating term of the Hamiltonian. The
resulting effective Hamiltonian can be written as (with $\hbar=1$)
\begin{equation}
H^{\prime}=\frac{1}{2}\left(
\begin{array}{ccc}
-\delta_{1} & 2\Omega & 0 \\
2\Omega & \delta_{1} & 2T_{e} \\
0 & 2T_{e} & \delta_{2}
\end{array}
\right), \label{t4}
\end{equation}
where $\delta_{1}=\omega_{01}-\omega$ is the detuning of the laser
with the exciton resonance, $\delta_{2}=2\omega_{12}+\delta_{1}$,
and $\omega_{ij}=(\varepsilon_{i}-\varepsilon_{j})$. Provided that
all remaining parameters change slowly, or assuming that the pulse
has a broad square shape, the time dependence of the effective
Hamiltonian can be neglected, and the state vector of the system
expressed as the superposition of the three eigenstates (or
dressed states \cite{Cohen92}) of Hamiltonian (\ref{t4}). This
problem has exact solution given by the roots of a cubic equation.
An interesting case is when the pulse is in resonance with the
exciton energy $\delta_{1}=0$ and the levels $|1\rangle$ and
$|2\rangle$ are also in resonance, resulting
$\delta_{2}=2\omega_{12}=0$ (this value of $\omega_{12}$ can be
tuned with an applied gate voltage). In this case the eigenvalues
of Hamiltonian (\ref{t4}) have the simple form, $\lambda_{0}=0$,
$\lambda_{\pm}=\pm \sqrt{\Omega^{2}+T_{e}^{2}}$ with corresponding
eigenstates
\begin{subequations}
\begin{eqnarray}
&&|\lambda_{0}\rangle=\cos\theta|0\rangle-\sin\theta|2\rangle, \\
&&|\lambda_{\pm}\rangle=\frac{1}{\sqrt{2}}\left(\sin\theta|0\rangle
\pm|1\rangle+\cos\theta|2\rangle\right), \label{t6}
\end{eqnarray}
\end{subequations}
where $ \cos \theta =T_{e}/\sqrt{\Omega^{2}+T_{e}^{2}}$.

\begin{figure}[tbp]
\includegraphics*[width=1.0\linewidth]{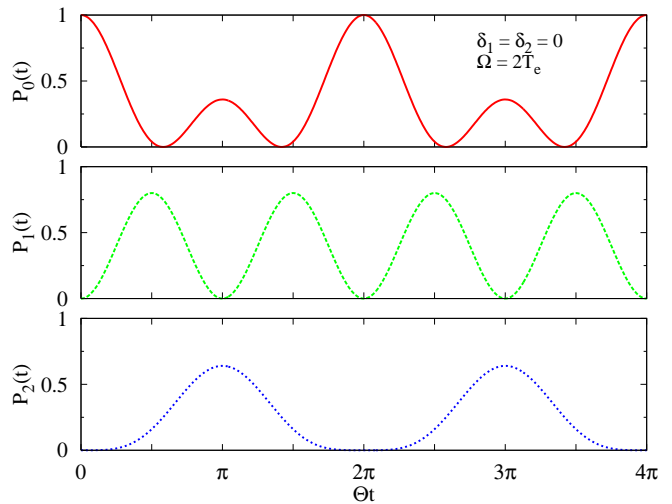}
\caption{(Color online) Population of the levels of the system as
given by Eqs.\ (\ref{t7}). Note that at $\Theta t =n\pi$, for $n$
odd, the system does not fully restore the population of
$|0\rangle$, as the system partially occupies state $|2\rangle$ if
$T_e\neq0$.} \label{prob}
\end{figure}

Assuming that we start the system in the ground state $|0\rangle$,
the occupation probability of the states of the Hamiltonian can be
expressed as
\begin{subequations}
\begin{eqnarray}
&&P_{0}(t)=\left|\sin^2\theta \cos(\Theta t)+\cos^2\theta\right|^{2}, \\
&&P_{1}(t)=\sin^2\theta\sin^{2}(\Theta t), \\
&&P_{2}(t)=\sin^2\theta \cos^2\theta \left|\cos(\Theta
t)-1\right|^{2},
\end{eqnarray} \label{t7}%
\end{subequations}
where $\Theta=\sqrt{\Omega^{2}+T_{e}^{2}}$. The result for
$\Omega=2T_e$ can be seen in Fig.\ \ref{prob}, where we notice
that the Rabi oscillations are incomplete, as interdot tunneling
transfers some of the population to the indirect exciton state
$|2\rangle$. Experimentally this could be very useful since one
could monitor the population of the second dot as a non-disturbing
probe of the coherent state of the QD. The presence of Rabi
oscillations generated by the optical pulse in the first dot could
then be directly measured. This can be done by sending a probe
pulse at the resonance frequency of the exciton in the second
(smaller) dot that has different (larger) frequency from the first
and would then measure the transient differential transmission,
reflecting its population. If the electron is in the small QD, the
resonant photon cannot be absorbed there, both because of Pauli
blocking of the electron, and because Coulomb blockade interaction
would require higher energy to create a charged exciton.
\cite{Gammon02} Another interesting possibility is to measure the
photocurrent induced by the pulse in a double QD diode structure,
similar to that used by Zrenner \textit{et al.} \cite{Zrenner02}
for a single dot. The photocurrent signal would be a direct
measure of how much the electron has tunnelled to the second dot.

\begin{figure}[tbp]
\includegraphics*[width=1.0\linewidth]{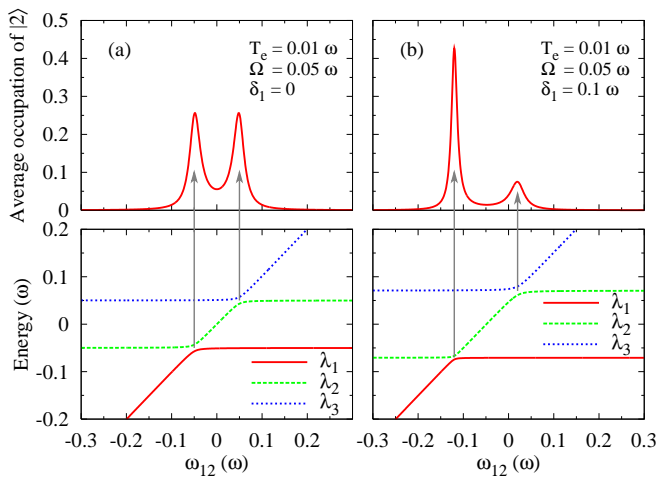}
\caption{(Color online) Average occupation of level $|2\rangle$ as
function of detuning between levels $|1\rangle$ and $|2\rangle$
($\omega_{12}$) for $T_e=0.01~\omega$ and $\Omega=0.05~\omega$. In
(a) the pulse is applied in resonance with the exciton energy
($\delta_1=0$) and (b) with a detuning of $\delta_1=0.1~\omega$.
In lower panels we show the respective dressed energy spectrum,
$\lambda_i$ eigenvalues of the effective Hamiltonian (\ref{t4}).
Arrows indicate that anticrossings in the spectrum yield an
efficient population transfer to level $|2\rangle$.}
\label{p2aver1}
\end{figure}

Although conceptually simple, the dynamics of the system presents
some surprises that are not intuitive.  For example, we would
expect that the best way to create the state with one electron in
the second dot, (state $|2\rangle$ in our model) would be to apply
a pulse in resonance with the exciton energy, so that we can
create one electron-hole pair in the first dot and then allow the
electron to tunnel to the second dot easily if its states are also
in resonance ($\epsilon_1=\epsilon_2$, $\omega_{12}=0$). This
expectation is in part wrong, as we can see in Fig.\
\ref{p2aver1}. We plot there the average occupation of level
$|2\rangle$ $\left[1/t_{\infty} \int_0^{t_{\infty}}
P_2(t)dt\right]$ as a function of the voltage-controlled detuning
of the levels $\epsilon_1$ and $\epsilon_2$ ($\omega_{12}$), for a
resonant pulse in (a) and out of resonance in (b), assuming
$\Omega=0.05\omega$ and $T_e=0.01\omega$. Notice that we have two
equal peaks in (a), but neither one is located when the levels are
in resonance ($\omega_{12}=0$). A better transfer can be reached
when {\em both}, the laser and the level detuning are different
from zero, as we can see in Fig.\ \ref{p2aver1}(b), where we have
a very narrow and high peak at $\omega_{12}/\omega \approx -0.12$.
To understand this behavior, we plot the energy spectrum as a
function of $\omega_{12}$ in the lower panels.  We can see that
the peaks in the average occupation of state $|2\rangle$ occur
exactly at the anticrossing positions in the dressed spectrum, as
indicated by arrows. An anticrossing indicates sizable mixing
between levels from the same subspace of a symmetry group in
Hilbert space. This mixing allow a maximum exchange of probability
between the states involved. In our case this will reflect a
maximum transfer of population to the level $|2\rangle$. Varying
the gate voltage in an asymmetric double dot diode to tune the
levels in and out of resonance would be an experimental
implementation of this idea, resulting in peaks in the
photocurrent at gate voltages satisfying the conditions above.
\footnote{We should mention that the confined Stark effect will
shift the energy levels in each dot in experiments, so that there
is a slight gate voltage dependence on $\omega_{01}(V)$, resulting
in a corresponding $\delta (V)$, easily compensated during system
characterization.}

\begin{figure}[tbp]
\includegraphics*[width=1.0\linewidth]{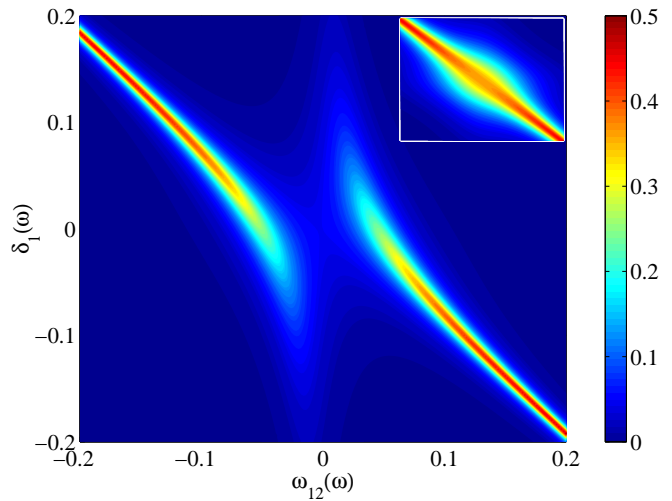}
\caption{(Color online) Average occupation of level $|2\rangle$ as
function of laser detuning ($\delta_1$) and level separation
($\omega_{12}$) for $T_e=0.01~\omega$ and $\Omega=0.05~\omega$.
Notice that if $\omega_{12}=0$ we have poor population transfer to
state $|2\rangle$. The best result is obtained when both, the
laser detuning ($\delta_1$) and the levels separation
($\omega_{12}$) are non-zero. Inset shows the result for
$T_e=\Omega=0.05~\omega$.} \label{p2aver2}
\end{figure}

Another experimental possibility is to keep the gate voltage fixed
(keep $\omega_{12}$ fixed) and vary the pulse frequency (vary the
detuning of the laser with the exciton energy $\delta_1$) while
measuring the induced photocurrent. The expected results can be
seen in Fig.\ \ref{p2aver2}. Notice that when the levels of the
different dots are in resonance ($\omega_{12}=0$) the effective
tunneling is weak for all range of detuning, re-enforcing the idea
discussed above about the relatively poor electron transfer when
the levels are in resonance.  In contrast, the result for
$\omega_{12}\neq0$, present a clear sharp peak. This highly
efficient transfer is likely to produce a clearly observable
result in experiments. Incidentally, average occupation for
$T_e\simeq\Omega$ results in strong transfer for
$\delta_1\simeq-\omega_{12}$ (inset, Fig.\ \ref{p2aver2}).

We should emphasize that these results represent nothing but the
control of the electron tunneling between dots using the coherent
Rabi oscillation induced by the laser in one of the dots and, in
this way, the manipulation of the states of the system. One could
imagine that such Rabi oscillation in an atomic molecule is also
possible, but the experimental ability to control the interdot
tunneling is unique to the QD system.  This control could have a
profound impact in the emergent field of quantum information
processing.

\begin{figure}[tbp]
\includegraphics*[width=1.0\linewidth]{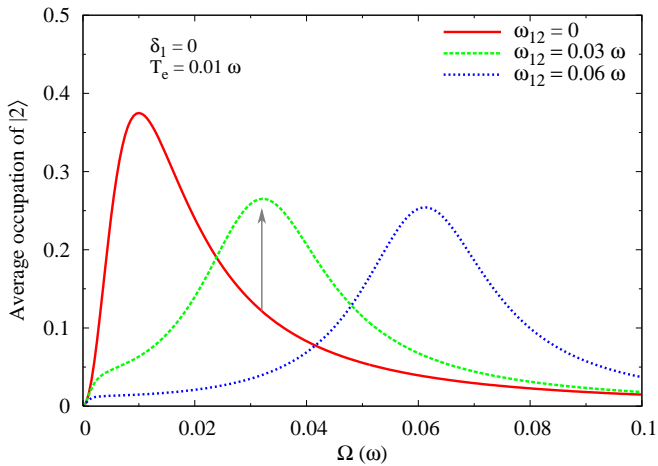}
\caption{(Color online) Average occupation of level $|2\rangle$ as
a function of pulse amplitude $\Omega$ for the laser frequency in
resonance with exciton energy ($\delta_1=0$), and $T_e=0.01
\omega$. Solid line is the result for $\omega_{12}=0$, dashed line
for $\omega_{12}=0.03\omega$ and dotted line for
$\omega_{12}=0.06\omega$. Arrow indicates that for $\Omega \approx
0.03\omega$, for example, there is an enhancement of the tunneling
probability compared with the on-resonant case ($\omega_{12}=0$).
Notice also tunneling is {\em suppressed} if pulse has high
amplitude.} \label{p2aver3}
\end{figure}

Another parameter also experimentally tunable is the amplitude (or
intensity) of the laser pulse, $\Omega$.  It is interesting to
note that for a given set of detuning and voltage values ($\delta$
and $\omega_{12}$), the average occupation of state $|2\rangle$ is
{\em not} a monotonic function of $\Omega$:  Figure \ref{p2aver3}
shows results for the average occupation of level $|2\rangle$ for
a resonant pulse ($\delta_1=0$) as a function of the pulse
amplitude $\Omega$ when levels $|1\rangle$ and $|2\rangle$ are in
and out of resonance. Note that when $\omega_{12}=0$, there is a
peak exactly at the point where $\Omega=T_e$, as one would expect
from a simple level mixing scheme, due to tunneling $T_e$
splitting the levels, and which are then effectively reconnected
by the pulse $\Omega$. If we further increase the amplitude, we
observe a {\em suppression} of the tunneling, where the average
occupation drops basically to zero. On the other hand, if the
levels are initially out of resonance, the tunneling is weak and
it can be substantially enhanced by increasing the intensity of
the pulse. Note for example, as indicated by the arrow in Fig.\
\ref{p2aver3}, that the tunneling for the out of resonance case is
higher than the case for resonant levels. It is clear that larger
level detuning $\omega_{12}$ requires a larger $\Omega$ to achieve
optimal transfer, but it is nevertheless always achievable, even
if the maximal amplitude is not as large as in the case
$\omega_{12}=0$. (This situation changes in fact, and if $\delta_1
\neq 0$, the maximum transfer to $|2\rangle$ occurs at finite
$\omega_{12}$ values -- not shown.)

We should be mindful that our solution here assumes constant
$\Omega$, and is therefore valid in the case of
slowly/smoothly-shaped pulses.  Shape forming and pulse sequence
design give additional flexibility to control the quantum
mechanical state of this QD molecule.  We will report elsewhere
our exploration of these degrees of freedom and the anticipated
advantages to control the system, useful perhaps in the nascent
field of quantum computing and QD optics.
\cite{Zrenner02,Gammon02}

We have studied a system of two coupled QDs, where the tunnel
coupling can be efficiently controlled and used to optically
monitor the Rabi oscillations in the system. The model can be
solved exactly for long constant-amplitude pulses. The results
show that we are able to control the tunneling by tuning the
parameters of the system such as the pulse intensity, laser
frequency, and gate voltage. Tunneling can be either enhanced or
suppressed, depending on the conditions. This provides an
electro-optical method to control the electron population of the
second dot. Experimentally one could monitor the population of the
second dot using a suitably tuned probe laser bean, which will
reflect the Rabi oscillations generated by the original optical
pump pulse. This opens the possibility to explore different
coherent states of coupled dot systems and allow their use in
novel quantum optics arrangements.

This work was partially supported by FAPESP and US DOE grant no.\
DE--FG02--91ER45334.  We thank C. J. Villas-B\^{o}as and N.
Studart for helpful discussions.

\end{document}